\documentclass{article}
\usepackage{dcase2023,amsmath,graphicx,url,times,booktabs,tabularx}
\usepackage{multirow}
\usepackage[inline]{enumitem}

\title{Crowdsourcing and Evaluating Text-Based Audio Retrieval Relevances \vspace{-12pt}}

\name{Huang Xie, Khazar Khorrami, Okko R\"as\"anen, Tuomas Virtanen \vspace{-12pt}}
\address{Unit of Computing Sciences, Tampere University, Finland \vspace{-18pt}}

\begin{document}
    \setlength{\abovedisplayskip}{3pt}
    \setlength{\belowdisplayskip}{3pt}

    \maketitle

    \begin{sloppy}

        \begin{abstract}
            This paper explores grading text-based audio retrieval relevances with crowdsourcing assessments.
            Given a free-form text (e.g., a caption) as a query, crowdworkers are asked to grade audio clips using numeric scores (between 0 and 100) to indicate their judgements of how much the sound content of an audio clip matches the text, where 0 indicates no content match at all and 100 indicates perfect content match.
            We integrate the crowdsourced relevances into training and evaluating text-based audio retrieval systems, and evaluate the effect of using them together with binary relevances arise from audio captioning.
            Conventionally, these binary relevances are defined by captioning-based audio-caption pairs, where being positive indicates that the caption describes the paired audio, and being negative applies to all other pairs.
            Experimental results indicate that there is no clear benefit from incorporating crowdsourced relevances alongside binary relevances when the crowdsourced relevances are binarized for contrastive learning.
            Conversely, the results suggest that using only binary relevances defined by captioning-based audio-caption pairs is sufficient for contrastive learning.
        \end{abstract}
        \vspace{-3pt}

        \vspace{-9pt}

        \section{Introduction}\label{sec:introduction}
        \vspace{-9pt}

        Text-based audio retrieval, or text-to-audio retrieval, refers to searching for audio clips with free text queries, which has great potential in real-world applications, such as search engines and multimedia databases.
        Early works~\cite{Slaney2002Semantic, Ikawa2018Acoustic} have mainly focused on methods of retrieving audio with carefully curated phrases (e.g., audio tags, onomatopoeic words).
        With the availability of large audio-caption datasets (e.g., Clotho~\cite{Drossos2020Clotho} and AudioCaps~\cite{Kim2019AudioCaps}) in recent years, increasing attention has been drawn to developing methods for audio retrieval using free-form text~\cite{Xie2022Language}.

        Most of the literature tackles text-to-audio retrieval with cross-modal learning methods.
        \mbox{Oncescu~\textit{et al.}~\cite{Oncescu2021Audio}} first established benchmarks in this topic with an adapted text-to-video retrieval model.
        With the recent success of large-scale pretrained audio models (e.g., PANNs~\cite{Kong2020PANNs}) and language models (e.g., BERT~\cite{Devlin2019BERT}), pretrained models are widely used for text-to-audio retrieval and fine-tuned on task-specific data to learn joint representations of audio and text~\cite{Xie2022Language}.
        Besides, several works~\cite{Mei2022On, Xie2023On} explored training strategies for text-to-audio retrieval models.
        For example,~\mbox{Mei~\textit{et al.}~\cite{Mei2022On}} evaluated several cross-modal learning objectives (e.g., InfoNCE loss~\cite{Oord2018Representation}) in the context of text-to-audio retrieval.

        In aforementioned works~\cite{Oncescu2021Audio, Mei2022On, Xie2023On}, relevances of audio to be retrieved are usually assumed to be binary, i.e., either relevant or irrelevant, given a text query.
        For example, an audio clip is relevant to a caption if and only if the caption thoroughly describes its sound content (i.e., their content matches perfectly).
        Practically, due to the lack of annotated non-binary relevances in existing datasets, binary relevances defined by captioning-based audio-caption pairs are adopted for system training and evaluation.
        Binary relevances are positive for audio-caption pairs where the caption describes the paired audio, and negative for all other pairs.
        This allows producing large quantities of positive and negative examples for contrastive learning in state-of-the-art systems~\cite{Xie2022Language}.

        As a caption might partially describe the sound content of an audio clip, we explore grading audio-text relevances with non-binary numerical scores.
        Specifically, we crowdsource audio-text relevances graded on a scale of 0 to 100, where 0 indicates completely irrelevant (i.e., no content match at all) and 100 indicates completely relevant (i.e., perfect content match).
        The main contributions of this work are:
        \begin{enumerate*}[label=\arabic*)]
            \item we crowdsource non-binary audio-text relevances for environmental audio and its existing captions;
            \item we integrate the crowdsourced relevances into training and evaluating text-to-audio retrieval systems, and evaluate the effect of using them alongside binary relevances defined by captioning-based audio-caption pairs;
            \item we release all the data and the crowdsourcing instructions to the research community to allow others to explore non-binary relevances~\cite{TAU2023GrRelDataset}.
        \end{enumerate*}
        \vspace{-12pt}

        \section{Crowdsourcing Assessments}\label{sec:crowdsourcing-assessments}
        \vspace{-9pt}

        This section presents the proposed method for crowdsourcing audio-text relevances.
        \vspace{-12pt}

        \subsection{Crowdsourcing Pipeline}\label{subsec:crowdsourcing-pipeline}
        \vspace{-6pt}

        We crowdsource audio-text relevances on Amazon Mechanical Turk (MTurk).
        Here we first introduce the terms used in this work.
        A~\textit{human intelligence task} (HIT) represents a single task that a crowdworker can work on.
        An~\textit{assignment} is a copy of a HIT that is assigned to a crowdworker.
        A worker~\textit{answer} is the submitted task result when a crowdworker completes an assignment.

        Fig.~\ref{fig:crowdsourcing_pipeline} presents an overview of the crowdsourcing pipeline.
        Audio clips and captions for relevance assessments are combined to form HITs, each of which consists of five audio clips and one caption.
        Every HIT is assigned to multiple MTurk workers.
        Raw answers containing graded audio-text relevances are collected and aggregated after workers complete their assignments.
        \vspace{-12pt}

        \begin{figure*}[!t]
            \centering
            \includegraphics[width=1.0\textwidth]{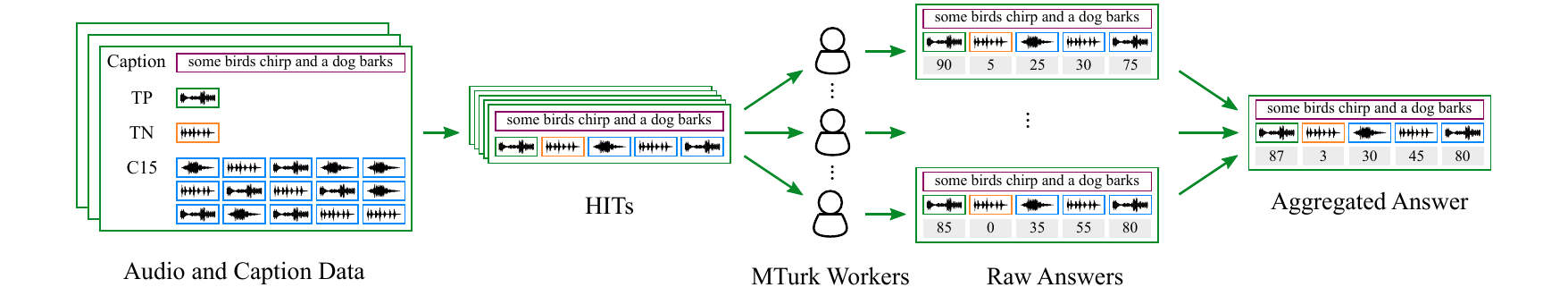}
            \vspace{-24pt}
            \caption{An overview of the pipeline for crowdsourcing relevance assessments.}
            \label{fig:crowdsourcing_pipeline}
            \vspace{-12pt}
        \end{figure*}

        \subsection{Audio and Caption Data}\label{subsec:audio-and-caption-data}
        \vspace{-6pt}

        We select a subset of captions and audio clips from each split in Clotho~\cite{Drossos2020Clotho}, with each subset containing 200 captions and 17 audio clips for each caption.
        Clotho crowdsources captions for each audio clip, and captions scored high by other workers results to the final captions~\cite{Drossos2020Clotho}.
        The audio clips selected for each caption consists of one true positive clip (TP, being completely relevant to the caption), one true negative clip (TN, being completely irrelevant to the caption), and 15 relevance-unknown candidates (C15).
        Specifically, we select audio clips corresponding to the captions in Clotho as TPs and obtain TNs using low audio-caption similarity scores estimated by the baseline system in DCASE 2023 Challenge Task 6B\footnote{https://dcase.community/challenge2023/task-language-based-audio-retrieval.} followed by human verification.
        For C15s, we choose the top five clips with high audio-caption similarity scores, together with another 10 randomly selected clips to include audio clips having diverse relevances.
        Table~\ref{tab:crowdsourcing_task_data_statistics} summarizes the captions and audio clips for crowdsourcing relevance assessments.
        \vspace{-12pt}

        \begin{table}[!t]
            \centering
            \begin{tabular}{c|c|c|c|c}
                \hline
                \multirow{2}{*}{\bfseries Split} & \multirow{2}{*}{\#\bfseries Captions} & \multicolumn{3}{c}{\#\bfseries Audio} \\
                \cline{3-5}
                &     & \#\bfseries TPs & \#\bfseries TNs & \#\bfseries C15s \\
                \hline
                development & 200 & 200             & 200             & 3000             \\
                \hline
                validation  & 200 & 200             & 200             & 3000             \\
                \hline
                evaluation  & 200 & 200             & 200             & 3000             \\
                \hline
            \end{tabular}
            \caption{Statistics of captions and audio clips selected for crowdsourcing relevance assessments.}
            \label{tab:crowdsourcing_task_data_statistics}
            \vspace{-12pt}
        \end{table}

        \subsection{Crowdsourcing Task Setting}\label{subsec:crowdsourcing-task-setting}
        \vspace{-6pt}

        For each caption, five HITs are created, each of which is assigned to several crowdworkers.
        Each HIT contains five audio clips, which are provided for crowdworkers to assess their individual relevance to the caption.
        The C15s of a caption are split into five batches of three C15s, with one batch per HIT.
        The TP and TN clips are used for quality check after crowdsourcing.

        In each assignment, crowdworkers are asked to assign numeric scores (between 0 and 100) to indicate their judgements of how much the sound content in each audio clip matches the given caption in that assignment.
        Inspired by~\cite{Roitero2021On}, we grade audio-text relevances on a scale of 0 to 100, where 0 indicates completely irrelevant (i.e., no content match at all) and 100 indicates completely relevant (i.e., perfect content match).
        An initial value of 0 is set as the default relevance score for each audio clip.
        Crowdworkers are required to listen to each audio clip entirely.
        \vspace{-12pt}

        \subsection{Quality Check}\label{subsec:quality-check}
        \vspace{-6pt}

        To collect high-quality answers, a quality check is conducted before and after crowdsourcing.
        Worker requirements are set up for selecting workers with high-quality work on MTurk (e.g., workers with a HIT approve rate greater than 98\%).
        Besides, crowdworkers should pass a predefined qualification test by correctly answering questions about identifying the audio clip described by a given caption from three candidates before they can accept our HITs (i.e., receiving assignments).

        With the fact that TPs are clearly more relevant than TNs within the same assignment, the former should receive higher scores than the latter.
        Let $s_{tp}$ be the graded relevance of a TP, and $s_{tn}$ be the one of the TN within the same assignment.
        Consistency verification on $s_{tp}$ and $s_{tn}$ is applied to check and select worker answers at the worker level.

        Let $D=\left\{ S_{i} \mid 1 \leq i \leq N \right\}$ denote the set of $N$ answers submitted by a worker, where $S_{i}=\left\{ s_{tp}^{i}, s_{tn}^{i}, s_{c1}^{i}, s_{c2}^{i}, s_{c3}^{i} \right\}$ represents the $i$-th answer from the worker, and $s_{c1}$, $s_{c2}$, $s_{c3}$ are the scores of the three C15 clips within the same assignment.
        As mentioned above, a C15 can be either completely relevant or irrelevant, or even partially relevant to a given caption.
        For every $S_{i} \in D$, we measure two random variables for the worker: $X$, which represents the difference of $s_{tp}$ and $s_{tn}$, and $Y$, which denotes the difference of every pair of $s_{c1}$, $s_{c2}$, $s_{c3}$.
        Intuitively, $s_{tp}$ should be higher than $s_{tn}$ by more than what is expected by chance when sampling from $Y$.
        For consistency verification on $s_{tp}$ and $s_{tn}$, we therefore require that $X$ and $Y$ should satisfy:
        \begin{equation}
            \label{eq:consistency_condition}
            E(X) = E(s_{tp} - s_{tn}) \geq E(Y) + \sigma(Y),
        \end{equation}
        where $E$ represents the expected value and $\sigma$ represents the standard deviation across $D$ (i.e., all answers from the worker).
        If the inequality is not satisfied, $D$ will be discarded completely.
        \vspace{-12pt}

        \section{Audio-Text Relevance Scores}\label{sec:audio-text-relevance-scores}
        \vspace{-9pt}

        This section analyzes the crowdsourced relevances.
        \vspace{-12pt}

        \subsection{Crowdsourced Raw Scores}\label{subsec:crowdsourced-raw-scores}
        \vspace{-6pt}

        Table~\ref{tab:crowdsourced_data} summarizes information about the crowdsourced data after filtering the data based on the quality check.
        For each HIT, answers were collected from at least five distinct crowdworkers.
        In total, 18204 answers were crowdsourced from 340 MTurk workers.

        \begin{table}[!t]
            \centering
            \begin{tabular}{c|c|c|c}
                \hline
                \bfseries Split & \#\bfseries HITs & \#\bfseries Workers & \#\bfseries Answers \\
                \hline
                development     & 1000             & 109                 & 6651                \\
                \hline
                validation      & 1000             & 113                 & 5064                \\
                \hline
                evaluation      & 1000             & 118                 & 6489                \\
                \hline
            \end{tabular}
            \caption{Statistics of crowdsourced data.}
            \label{tab:crowdsourced_data}
            \vspace{-12pt}
        \end{table}

        Fig.~\ref{fig:crowdsourced_raw_scores} presents the distribution of raw relevance scores of TP, TN, and C15 clips.
        For TPs, approximately 60\% of relevance scores have a value of 100.
        For TNs, about 90\% of relevance scores are zeros, and over 98\% of these scores are less than 20.
        It indicates that most crowdworkers can appropriately assess the relevances of TPs and TNs to a given caption.
        For C15s, over 10\% of relevance scores have a value of 100, which indicates that some C15s are highly relevant to a given caption.
        We notice that around 20\% of relevance scores of TPs are zeros, which shows the necessity of further processing on the crowdsourced raw scores.
        \vspace{-12pt}

        \begin{figure*}[!t]
            \centering
            \includegraphics[width=1.0\textwidth]{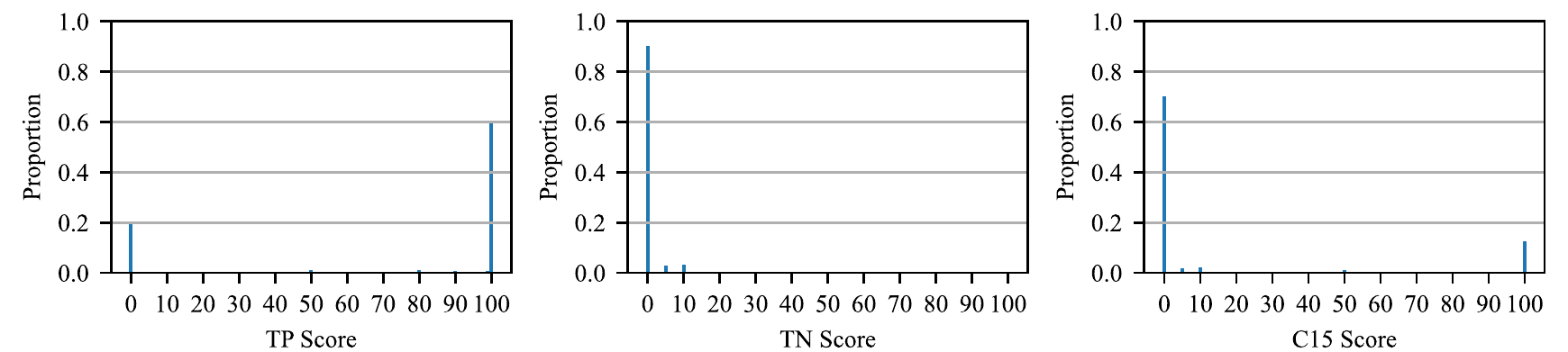}
            \vspace{-24pt}
            \caption{Distribution of raw relevance scores of TP, TN, and C15 clips.}
            \label{fig:crowdsourced_raw_scores}
            \vspace{-9pt}
        \end{figure*}

        \subsection{Aggregated Scores}\label{subsec:aggregated-scores}
        \vspace{-6pt}

        The raw scores from different workers regarding the relevance of an audio clip to a text query are aggregated by discarding a maximum and a minimum score and then averaging the remaining to produce a statistic that is robust to outliers.
        Fig.~\ref{fig:crowdsourced_aggregated_scores} presents the distribution of aggregated relevance scores of TP, TN, and C15 clips.
        After aggregating, the distribution of relevance scores becomes more balanced, with fewer instances of extreme or polarized judgements (e.g., scores of 0 and 100).
        Particularly, TPs exhibit a broader spectrum of relevances (e.g., having scores spanning from 30 to 100) compared to other clips.
        Over 99\% of TPs have a score above 10, while about 99\% of TNs have a score below 10.
        Around 30\% of C15s have a score above 10, and roughly 10\% have a score above the mean score of TPs (i.e., $s>72$).
        \vspace{-12pt}

        \begin{figure*}[!t]
            \centering
            \includegraphics[width=1.0\textwidth]{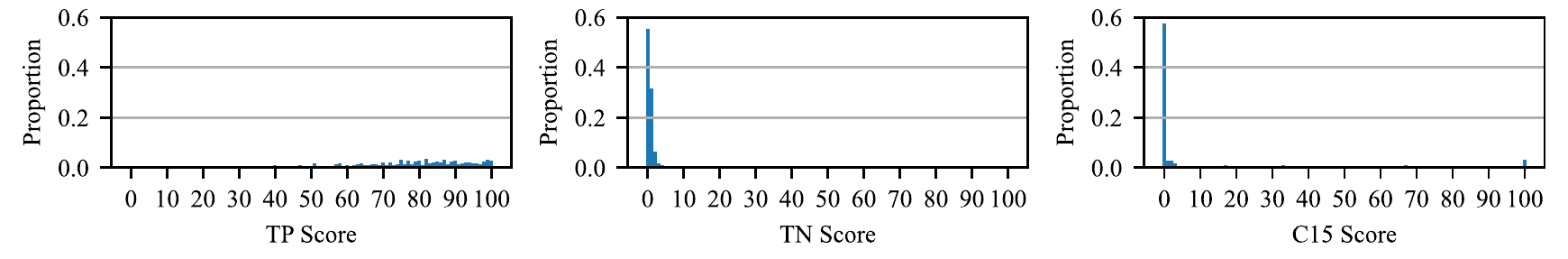}
            \vspace{-24pt}
            \caption{Distribution of aggregated relevance scores of TP, TN, and C15 clips.}
            \label{fig:crowdsourced_aggregated_scores}
            \vspace{-9pt}
        \end{figure*}

        \section{Experiments}\label{sec:experiments}
        \vspace{-9pt}

        This section reports experimental results of using the crowdsourced relevances for text-to-audio retrieval.
        Due to the lack of established methods for using non-binary relevances for training and evaluation, we binarize the crowdsourced relevances.
        \vspace{-12pt}

        \subsection{Audio-Caption Pairs}\label{subsec:audio-caption-pairs}
        \vspace{-6pt}

        Similar to previous studies~\cite{Xie2022Language}, we tackle text-to-audio retrieval with contrastive learning.
        To obtain positive and negative examples for contrastive learning, we binarize the crowdsourced relevances (see~\ref{subsec:aggregated-scores}) using the mean score of TPs as an arbitrary threshold.
        Specifically, we obtain positive audio-caption pairs by combining:
        \begin{enumerate*}[label=\arabic*)]
            \item a caption with its high-graded C15 clips, which have a score above the threshold;
            \item the TP clip of a caption with the captions corresponding to its high-graded C15 clips in Clotho;
            \item the siblings (i.e., captions describing the same TP clip in Clotho) of a caption with its high-graded C15 clips.
        \end{enumerate*}
        All other audio-caption combinations are treated as negative pairs.
        The resulting positive and negative pairs are referred to as content-matching pairs with ``Binarized Crowdsourced Relevances'' (BiCrRel).

        As a baseline, we created a subset of Clotho by selecting those audio-caption pairs of which the audio or the caption were part of BiCrRel, i.e., using the captioning-based clip-specific audio-caption pairs from Clotho~\cite{Drossos2020Clotho}.
        The selected pairs are referred to as captioning-based pairs with ``Binary Relevances'' (BiRel), which include the same audio clips and captions as in BiCrRel.
        Table~\ref{tab:positive_pairs} summarizes information about BiCrRel, BiRel, and their combination (``BiCrRel+BiRel'').
        The development / validation / evaluation splits are used for training / validation / evaluation, respectively.
        \vspace{-12pt}

        \begin{table}[!t]
            \centering
            \begin{tabular}{c|c|c|c}
                \hline
                \bfseries Split & \bfseries BiCrRel & \bfseries BiRel & \bfseries BiCrRel+BiRel \\
                \hline
                development     & 3890              & 2370            & 6260                    \\
                \hline
                validation      & 2560              & 1580            & 4140                    \\
                \hline
                evaluation      & 2440              & 1390            & 3830                    \\
                \hline
            \end{tabular}
            \caption{Number of positive audio-caption pairs in BiCrRel, BiRel, and their combination (``BiCrRel+BiRel'').}
            \label{tab:positive_pairs}
            \vspace{-12pt}
        \end{table}

        \subsection{Retrieval System}\label{subsec:retrieval-system}
        \vspace{-6pt}

        We experiment with the retrieval system proposed as the baseline in DCASE 2023 Challenge Task 6B\footnote{https://dcase.community/challenge2023/task-language-based-audio-retrieval.}, where a pretrained CNN14~\cite{Kong2020PANNs} is employed as the audio encoder and the Sentence-BERT (i.e., ``all-mpnet-base-v2'')~\cite{Reimers2019Sentence} is used as the text encoder.
        This system is trained by optimizing the InfoNCE loss~\cite{Oord2018Representation} such that embeddings of the paired audio and text are pulled together while those of the unpaired are pushed far away.

        \textbf{Audio Encoder}.
        The CNN14~\cite{Kong2020PANNs}, which is pretrained on AudioSet~\cite{Gemmeke2017AudioSet}, is employed as the audio encoder, with its last linear layer discarded.
        An extra linear layer is added on the top to generate 300-dimensional audio embeddings.
        The audio encoder is fine-tuned during training.

        \textbf{Text Encoder}.
        The Sentence-BERT~\cite{Reimers2019Sentence}, which is derived from BERT~\cite{Devlin2019BERT} for the purpose of generating robust sentence embeddings, is used as the text encoder.
        An extra linear layer is also added on the top to generate 300-dimensional text embeddings.
        The Sentence-BERT is frozen during training.

        \textbf{InfoNCE Loss}.
        The InfoNCE loss~\cite{Oord2018Representation} is a symmetric cross-entropy loss, taking the form of
        \begin{equation}
            \label{eq:infonce_loss}
            \begin{split}
                \mathcal{L} = - \dfrac{1}{M} \sum_{i=1}^{M} [ \log \dfrac{\exp(z_{ii} / \tau)}{\sum_{j=1}^{M} \exp(z_{ij} / \tau)} + \log \dfrac{\exp(z_{ii} / \tau)}{\sum_{j=1}^{M} \exp(z_{ji} / \tau)} ],
            \end{split}
        \end{equation}
        where $\tau$ represents the temperature hyper-parameter, $M$ denotes the number of audio-text pairs, and $z_{ij}$ represents the cosine similarity of the $i$-th audio embedding and the $j$-th text embedding.
        It has been widely used to train cross-modal retrieval systems~\cite{Xie2022Language}.

        \textbf{Training Setup}.
        The retrieval system is trained with mini-batches consisting of 32 audio-text pairs from a development split.
        An Adam optimizer with an initial learning rate of $0.001$ is adopted to optimize training.
        Learning rate is reduced by a factor of ten once the validation loss does not improve for five epochs.
        Training is terminated by early stopping with a patience of ten epochs.
        \vspace{-12pt}

        \subsection{Evaluation Metrics}\label{subsec:evaluation-metrics}
        \vspace{-6pt}

        Retrieval performance is measured with recall at 10 (R@10) on different evaluation splits.
        The R@10 is defined as the proportion of relevant items among the top 10 results to all the relevant items contained in the data and is averaged over queries~\cite{Xie2022Language}.
        The more relevant items are within top 10 results, the higher R@10 it is.
        \vspace{-12pt}

        \subsection{Results}\label{subsec:results}
        \vspace{-6pt}

        Table~\ref{tab:experimental_results} shows that training the system on BiRel leads to high R@10 (e.g., 0.566 on BiRel), whereas training the system on BiCrRel yields low R@10 (e.g., 0.478 on BiRel) and on BiCrRel+BiRel results in intermediate R@10 (e.g., 0.509 on BiRel).
        We conclude that the crowdsourced relevances do not improve the performance from captioning-based audio-caption pairs when they are reduced to binary relevances.
        A possible explanation is that captions in Clotho are crowdsourced specifically to describe an exact audio clip, while the crowdsourced relevances are graded based on their matching content with a given caption (i.e., different underlying purposes and criteria for generating captions and assessing relevances).
        Besides, with the fact that each caption in BiRel is annotated with one relevant audio clip while a caption in BiCrRel can have several relevant audio clips, it makes text-to-audio retrieval on BiCrRel more difficult and leads to a decrease in R@10.
        \vspace{-12pt}

        \begin{table}[!t]
            \centering
            \begin{tabular}{r|c|c|c}
                \hline
                \multirow{2}{*}{\bfseries Training Data} & \multicolumn{3}{c}{\bfseries Evaluation Data} \\
                \cline{2-4}
                & \bfseries BiCrRel & \bfseries BiRel & \bfseries BiCrRel+BiRel \\
                \hline
                BiCrRel       & 0.357             & 0.478           & 0.407                   \\
                \hline
                BiRel         & 0.412             & 0.566           & 0.479                   \\
                \hline
                BiCrRel+BiRel & 0.363             & 0.509           & 0.426                   \\
                \hline
            \end{tabular}
            \caption{Evaluation R@10 of text-to-audio retrieval with the retrieval system trained on different pairs.}
            \label{tab:experimental_results}
            \vspace{-12pt}
        \end{table}

        \section{Conclusions}\label{sec:conclusions}
        \vspace{-9pt}

        We explore grading audio-text relevance for text-based audio retrieval via crowdsourcing assessments.
        We crowdsource audio-text relevances graded on a scale of 0 to 100, where 0 indicates completely irrelevant and 100 indicates completely relevant.
        We integrate crowdsourced relevances into training and evaluating text-to-audio retrieval systems, and evaluate the effect of using them alongside binary relevances defined by captioning-based audio-caption pairs.
        Experimental results show that the crowdsourced relevances do not positively contribute to the performance when they are reduced to binary relevances, and using only binary relevances defined by captioning-based audio-caption pairs is sufficient for contrastive learning.
        \vspace{-12pt}

        \section{Acknowledgment}\label{sec:acknowledgment}
        \vspace{-9pt}

        The research leading to these results has received funding from Emil Aaltonen foundation funded project ``Using language to interpret unstructured data'' and Academy of Finland grant no. 314602.

        \bibliographystyle{IEEEtran}
        \bibliography{refs}

%
%
%
%
%
%
%
%
%

    \end{sloppy}
\end{document}